\documentclass[article]{has40}
\usepackage{graphicx}
\usepackage{amssymb}
\tabletypesize{\normalsize}  

\def\plotone#1{\centering \leavevmode
\includegraphics[width=.95\columnwidth]{#1}}

\def\plotone#1{\centering \leavevmode
\includegraphics[width=.95\columnwidth]{#1}}

\shortauthors{Kinemuchi}
\shorttitle{KIC 9832227}

\begin{document}
\large    
\pagenumbering{arabic}
\setcounter{page}{92}

\title{To Pulsate or to Eclipse?  Status of KIC 9832227 variable star}

%
%
\author{{\noindent Karen Kinemuchi{$^{\rm 1}$}\\
\\
{\it (1) Apache Point Observatory/New Mexico State University}
}}

%
%
\email{(1) kinemuchi@apo.nmsu.edu}


\begin{abstract}
The variable star known as NSVS 5597754 was identified as a short
period RRc field star from ground-based photometry.  This star happens
to fall in the Kepler field of view and has been observed for 3 years
with the Kepler spacecraft.  Renamed KIC 9832227, 
this star was reidentified as an eclipsing binary with a periodicity
of 0.457970 days.  The NSVS ground based photometry yielded a period
solution of 0.229 days.  The Kepler photometry, which sampled the
light cycle every 30 minutes, shows a peculiar behavior that is not
purely eclipsing superposed over the primary variation.  Follow-up ground based observations were taken in
2010 with the 0.6m MSU Observatory to obtain additional information.
I present a summary of the ground based data taken of this perplexing
star and conclusions derived thus far. 
\end{abstract}

\section{Introduction}
The Kepler spacecraft has revolutionized stellar pulsation and
asteroseismological studies.  Many well known variable stars are now
cast in a new light from the monitoring performed by the spacecraft.  An example of new discoveries made with Kepler in the realm
of RR Lyrae variable stars is the period doubling found in some
Blazhko effect affected stars (Szabo et al. 2010).  Although a
relatively small number of RR Lyrae variable stars are within the
Kepler field, each star holds a wealth of information, and puzzles, to
keep astronomers busy.

I present one such variable star which was previously identified as an
RR Lyrae star, but Kepler has reclassified it and uncovered peculiar
behavior never seen from ground based data.  This star has multiple
nomenclature, but will be mostly referred to by the Kepler Input
Catalog designation, KIC 9832227.  The position of this star is at
$\alpha_{2000.0} = 19^{h} 29' 15.95"$ and $\delta_{2000.0} =
+46^{\circ} 37' 19.88"$, and it has $V$ magnitude of 12.369 (Pigulski et
al. 2009).  I discuss the analysis performed currently on the ground
based data, and suggest possible explanations of the peculiar behavior seen from
the Kepler light curves.  

\section{Past ground based observations}
\begin{figure*}
\centering
\plotone{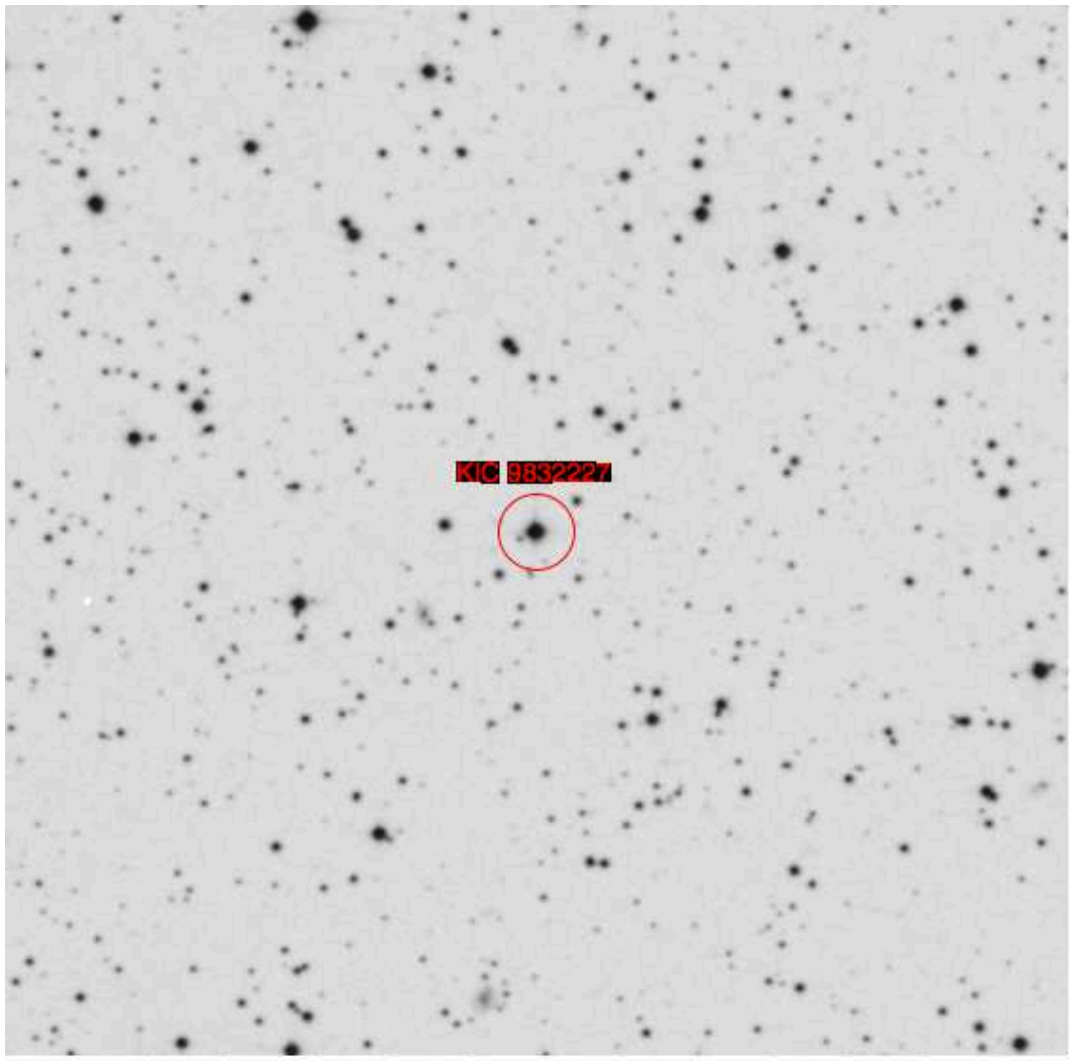}
\vskip0pt
\caption{Digital Sky Survey image of NSVS 5597754/KIC 9832227.  North
  is up and East is to the right.  The image covers $\sim 1$ arcminute
in right ascension and $\sim 10$ arcminutes in declination.}
\label{ds9pic}
\end{figure*}

\begin{figure}
\centering
\plotone{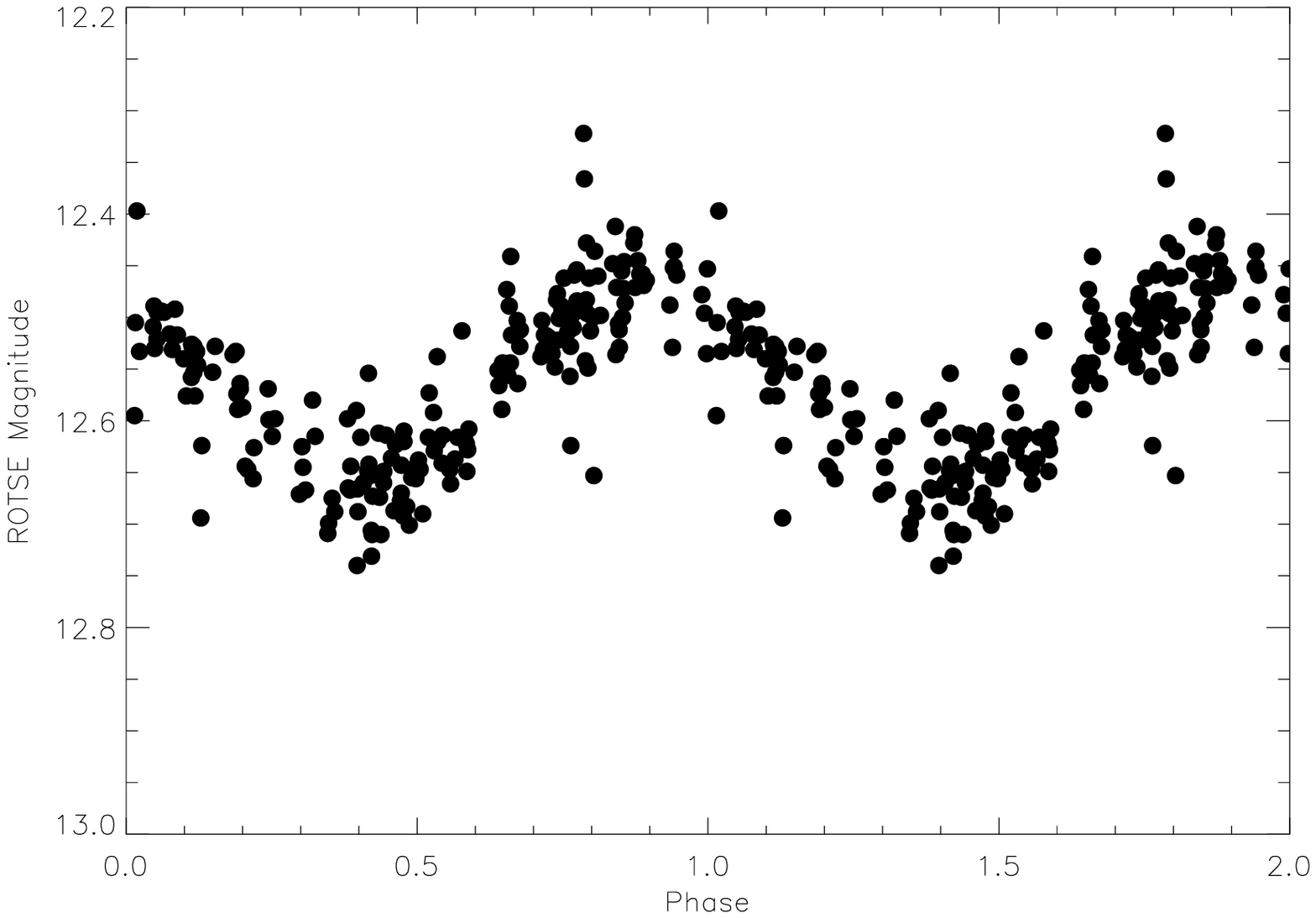}
\vskip0pt
\caption{Phased light curve for NSVS 5597754.  The photometry is based
on the original ROTSE magnitude system which is white light.}
\label{nsvsltc}
\end{figure}

The variable star seen in Figure \ref{ds9pic} was identified in an all-sky
survey (Northern Sky Variability Survey: NSVS) completed with a
ground-based, robotic telescope, ROTSE-I (Robotic Optical Transient
Search Experiment) (Kinemuchi et al. 2006; Wozniak et al. 2004).  Over 300 observations were taken with
the robotic telescope over a course of a year.  The cadence of
observations was two visits per night while that portion of the sky
was visible from northern New Mexico. The original robotic telescope
collected data without a filter, and hence, the measured magnitudes
are in white light.  Figure \ref{nsvsltc} shows the phased light curve of the photometry performed on
the star.  Kinemuchi et al. 2006 identified this star, dubbed NSVS
5597754, as an RR Lyrae variable star, Bailey type c.   The
classification was based upon the period solution ($P=0.229$ days) and
the sinusoidal light curve shape.

This variable star was also observed in the All-Sky Automated
Survey (ASAS: Pigulski et al. 2009) in the $V$ and $I$-bandpasses.
From their observations, this star has a $V-I$ color of 0.719  ($V=12.369$
and $I= 11.650$) and a $V$ amplitude of 0.22 mag.  The star was
reported to be an eclipsing binary star (W Ursa Majoris type or EW)
with a period of 0.4579480 days.

As demonstrated by the ground based data, the period solution depends
upon the interpretation of the phased light curve.  The shorter period
of 0.229 days appears consistently if one assumes the star is
pulsational in nature (see Section 4).  The longer period of 0.457
days is derived when the phased light curve shows an eclipsing binary
star.  Period aliases no longer become an obstacle with the near
constant monitoring performed by the Kepler spacecraft, as discussed
in the following section.

\section{Kepler observations}
 The Kepler spacecraft, launched in 2009, has been observing KIC
9832227 since the beginning of the mission, and has a total coverage
of over 1300 days.  KIC 9832227 has been observed in long cadence
mode, meaning observations were taken every 29.4 minutes.  The exquisite
coverage shows detailed luminosity changes in this star that have not
been discovered with the ground based data thus far.

\begin{figure*}
\centering
\includegraphics[scale=0.6,angle=-90]{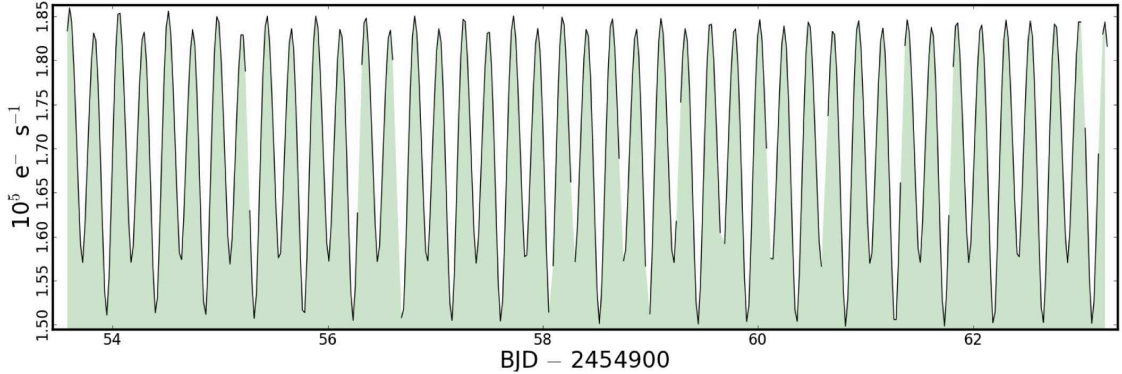}
\caption{Kepler light curve of KIC 9832227 during Quarter 0.
  Observations were taken in long cadence mode (every 29.4 minutes).
  The y-axis is the Pre-Data Conditioned Simple Aperture Photometry
  Flux in the units of electrons per second.  The x-axis is the
  Barycentric Julian Date in days.}
\label{q0}
\end{figure*}

\begin{figure*}
\centering
\includegraphics[scale=0.6,angle=-90]{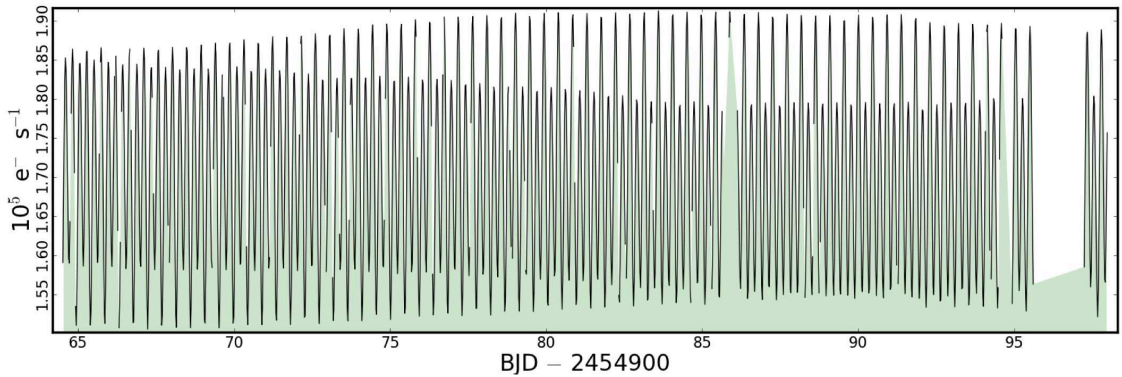}
\vskip0pt
\caption{Kepler light curve of KIC 9832227 during Quarter 1  (34 days).  Data
  were collected in long cadence mode.  The units on the x and y-axis
  are the same as in Figure \ref{q0}.}
\label{q1}
\end{figure*}

\begin{figure*}
\centering
\includegraphics[scale=0.6,angle=-90]{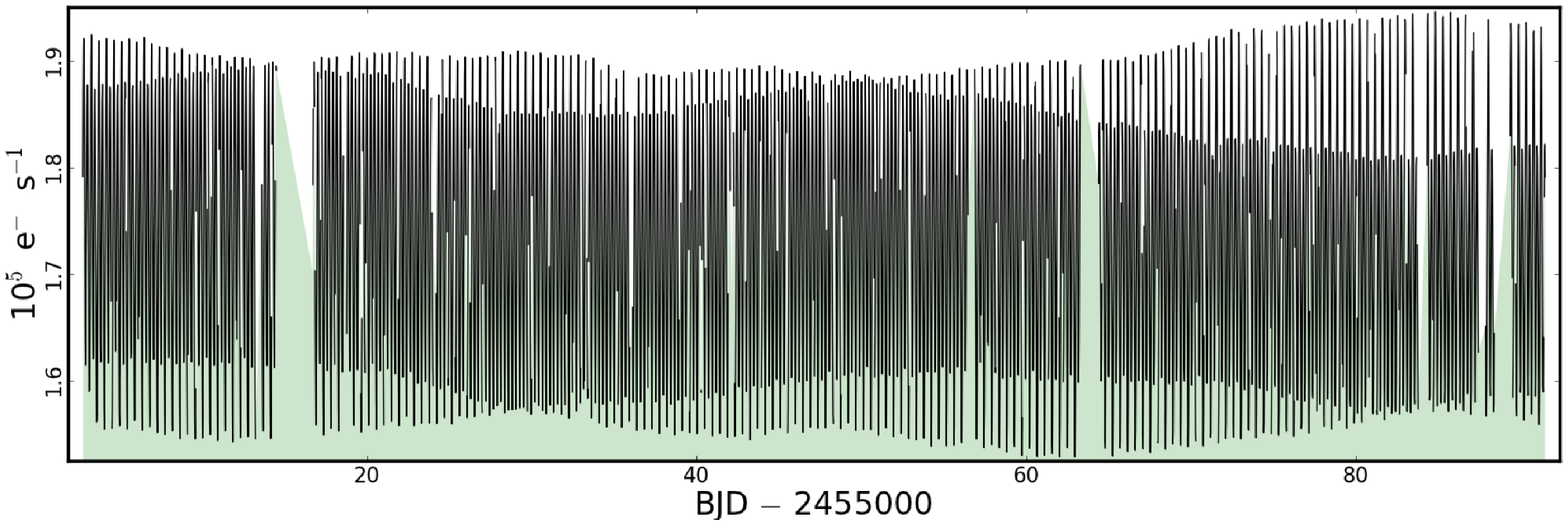}
\vskip0pt
\caption{Kepler light curve of KIC 9832227 during Quarter 2 (93
  days). The units on the x and y-axis
  are the same as in Figure \ref{q0}. }
\label{q2}
\end{figure*}

\begin{figure*}
\centering
\includegraphics[scale=0.6,angle=-90]{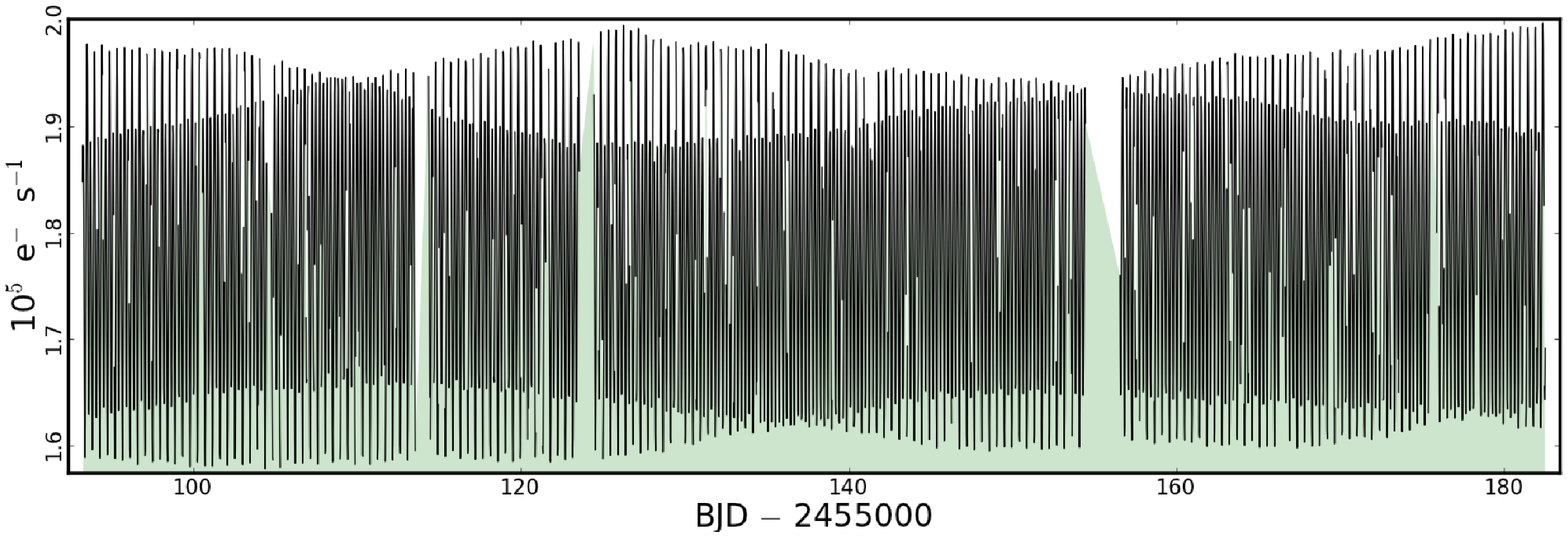}
\vskip0pt
\caption{Kepler light curve of KIC 9832227 during Quarter 3.  The units on the x and y-axis
  are the same as in Figure \ref{q0}.}
\label{q3}
\end{figure*}

During the commissioning period, known as Quarter 0 (Q0), this star was
observed for 10 days at long cadence mode.  The light curve from this
time period shows a clear eclipsing nature (see Figure \ref{q0}).  The star
was observed during Quarter 1 (Q1), which lasted 34 days, and a 
curious behavior appeared that was not seen in the 10 day coverage of
Q0. A secondary variation can be clearly seen over the primary
variation.  This ``envelope'' variation continues into subsequent
quarters.  Figure \ref{q1} - \ref{q3} shows Q1, Quarter 2 (Q2), and Quarter
3 (Q3) data of KIC 9832227 and an overall view of the secondary
variation.  Note that Q2 and Q3 have 93 days of coverage per quarter.

The Kepler Eclipsing Binary Working Group analyzed the data and
determined this star to be an ``uncertain''
classification\footnote{http://keplerebs.villanova.edu/} (Matijevi{\v c}
et al. 2012).  They determined the variation period to be 0.457950 days, which is nearly double
the period value determined from the NSVS photometry.  

A recent update of the stellar parameters in the Kepler Input
Catalog (Pinsonneault et al. 2012) shows that KIC 9832227 had an
effective temperature of 5854K and a log $g$ value of 4.45.  The
effective temperature indicates this star is too cool to be a c-type
RR Lyrae, which typically is much warmer as the blue edge of the
instability strip has an effective temperature of 7400K (Smith
1994).  The log $g$ value also shows that this star is a dwarf.  With
the addition of the stellar parameter information, we can conclusively
say that this star is not an RR Lyrae star. 

\begin{figure*}
\centering
\includegraphics[angle=90,scale=0.6]{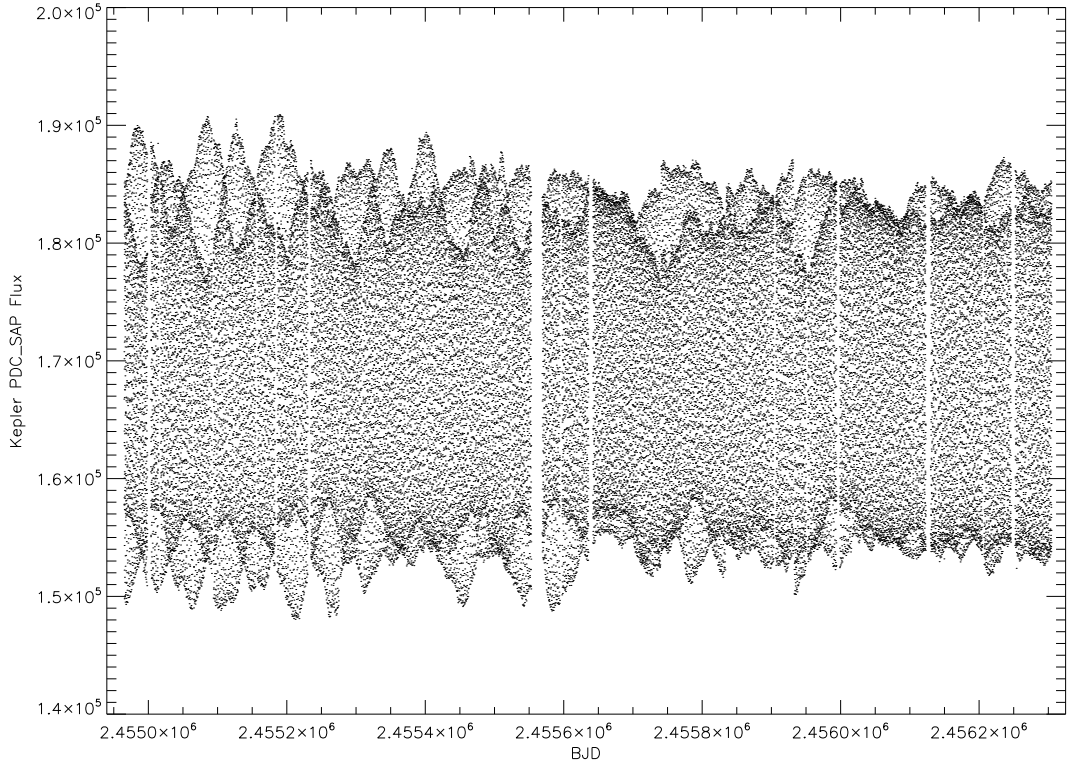}
\vskip0pt
\caption{Kepler light curve of KIC 9832227 from Q1-Q15.  Data were
  collected in long cadence mode.}
\label{allq}
\end{figure*}

The entire data set collected for this star are currently
available through the Mikulski Archive for Space
Telescopes\footnote{http://archive.stsci.edu/kepler/}.  Figure
\ref{allq} shows the light curve from 2009 to 2012 (Q0-Q15).  The data from each quarter were
offset by a mean amount in flux units with respect to Quarter 5 data
to stitch multiple quarters together.  No scaling correction was applied to the quarter
datasets.  Using the first 500 Kepler observations, we recalculated
the primary periodicity using the Supersmoother algorithm (Reimann
1994), and we arrive at $P=0.45797$ days for an eclipsing binary.
This solution agrees with the findings of the Eclipsing Binary Working
Group.  We note the long term ``envelope'' variation seen over the primary is not periodic, and
at Quarter 13, it appears to disappear for about 36 days.   The nature of this
envelope feature is currently being investigated.  Suggestions for
this phenomenon are discussed in Section 5.

\section{Follow-up Photometry and Analysis}
\begin{figure*}
\centering
\includegraphics[width=12cm,angle=-90]{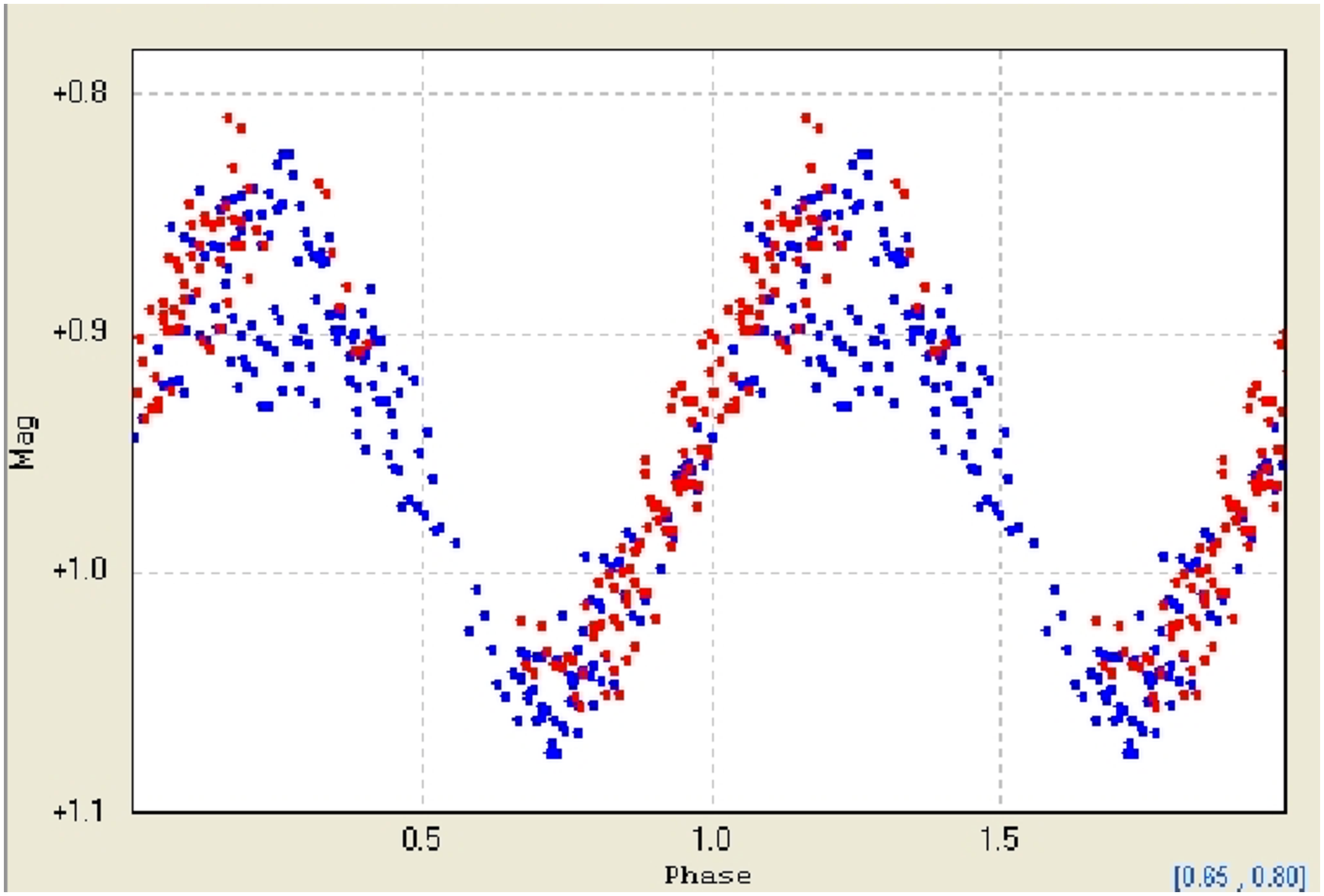}
\vskip0pt
\caption{Ground-based photometry of KIC 9832227 from the Michigan
  State University Observatory 0.6m telescope.  Data were collected by
H.A. Smith during the fall of 2010 (red points) and summer of 2011
(blue points).}
\label{msuphot}
\end{figure*}

During Quarter 7 and 10 Kepler monitoring, follow-up ground based observations of
this star were collected at the Michigan State University 0.6m
telescope in multiple bandpasses (Johnson $BVI$ and Sloan $g'r'i'$
filters).  Figure \ref{msuphot} shows the phase-folded light curve from the MSU
observations.  We see a notable ``peakiness'' at the maximum and
minimum of the variation rather than a more rounded feature.  A
periodicity of 0.22898 days was determined for these datasets, but it
is clear that with this period the light curve did not repeat
precisely from cycle-to-cycle.

Further period analysis was done on the Kepler photometry from
Q0-Q5.  The light curve was synthesized by R. Stellingwerf with Period Dispersion
Minimization (PDM) (Stellingwerf 1978).  A primary period of 0.22898
days was also found, and a secondary period of 0.4601 days was
discovered.  The secondary variation was modelled, but the results
yielded no clear picture as to what the phenomenon is.

Under the premise that this star had pulsations like a c-type RR
Lyrae star, Fourier Decomposition techniques were applied to both NSVS
and Kepler data.  This work was completed by S. Morgan at University
of Northern Iowa.  In Table 1, we list the resultant Fourier
decomposition parameters for the light curves developed from the NSVS
data (300 epochs) and Kepler data (23000 observations).  We list only
the Fourier decomposition parameters for $n=2$ and $n=3$.  The NSVS
data produced a period solution of 0.228958 days, and the Kepler data
had a period of 0.228975 days.  From the analysis of the Kepler data,
a secondary period of 0.4598 days appeared, but has a power 30 times
smaller than the primary period.  The discrepancy on the Fourier
parameters from the two datasets could come from analyzing light
curves from a non-standard bandpass (i.e. usually Fourier
decomposition analysis is done on broad-band Johnson $V$ data).
The NSVS dataset was obtained from white light, and the Kepler
data were taken from a broad optical filter ($\lambda = 4300-9000
\AA$).  The Fourier decomposition parameter discrepancy between data
sets could also be explained that with the Kepler data, the binarity
is more clear than any pulsational feature.

\begin{deluxetable}{cccccc}
\tablecaption{Fourier decomposition parameters.  Analysis done by S. Morgan}
\tablenum{1}
\tablewidth{0pc}
\tablehead{\colhead{Dataset} & \colhead{Period (days)} & \colhead{$R_{21}$} & \colhead{$\phi_{21}$} & \colhead{$R_{31}$} & \colhead{$\phi_{31}$} }

\startdata
NSVS & 0.228958 & 0.130 & 6.113 & 0.0366 & 1.857 \\
Kepler & 0.228975 & 0.081 & 3.118 & 0.030 & 6.245 \\
\enddata
\end{deluxetable}

\section{Conclusions}
KIC 9832227, or NSVS 5597754, is not a short period c-type RR Lyrae
star, but most likely a complicated W Ursa Majoris type eclipsing
binary.  The photometry, coupled with the derived stellar parameters
from Pinsonneault et al. 2012, shows this star is too cool to be an RR Lyrae,
and is characterized as a dwarf star from its log $g$ value.  Ground
based observations do not show the secondary variation that appears on
top of the primary 0.456 day periodicity.  Only through Kepler long
cadence observations were we able to see this additional behavior.

As to what could cause the secondary variation, a couple of
suggestions have been proposed.  If we use the period solution from
the PDM and Fourier decomposition of the light curve, the periodicity
is on the short end of the RR Lyrae regime.  If this star was a true
pulsator, it could be reclassified as an unusual delta Scuti variable
star.  Another explanation is that this star is an exotic binary star
system.  The secondary variation is not completely periodic, and in
the case of Quarter 13, the variation disappears for approximately 36 days.

Combining additional spectroscopic information with the accumulated
photometry can help illuminate the nature of this object.  Since
August 2012, I have collected 6 epochs worth of echelle spectra on
this object at the ARC 3.5m telescope in an effort to obtain radial
velocities and metal abundances.  With these additional pieces of
information, we hope to get a better idea of the composition of the
primary star and the dynamics of the system. 

\section{Acknowledgements}
I would like to thank Horace A. Smith for being my Ph.D
advisor at Michigan State University.  He originally
brought the NSVS project to my attention as a possible thesis project.  It is through that work I first discovered this
variable star, and while it is not an RR Lyrae, it has proven to be a
very interesting object.  

I would also like to thank all the collaborators who have provided
very useful discussions and head scratching over this star:  Robert
Stellingwerf for doing a careful period analysis of five quarters of
Kepler data; Siobahn Morgan for her Fourier decomposition work on both
the NSVS and Kepler data; Doug Welch, Peter Stetson,
and Michel Breger for delightful conversation and musing over this
peculiar star. 

Kepler data collected during Quarters 10,11,12 and 13 was made
possible via the Kepler Guest Observer Cycle 3 Program, ID\# 30024.
This paper includes data collected by the Kepler mission. Funding for
the Kepler mission is provided by the NASA Science Mission
directorate.  Some/all of the data presented in this paper were
obtained from the Mikulski Archive for Space Telescopes (MAST). STScI
is operated by the Association of Universities for Research in
Astronomy, Inc., under NASA contract NAS5-26555. Support for MAST for
non-HST data is provided by the NASA Office of Space Science via grant
NNX09AF08G and by other grants and contracts. 


\end{document}